\documentclass{article}
\usepackage{spconf,amsmath,graphicx,hyperref}
\usepackage{amsfonts}
\usepackage{xcolor}
\usepackage{booktabs}
\usepackage{multirow}
\usepackage{enumitem}

\hypersetup{hidelinks}

\title{Expressive Guitar Synthesis via Controllable Latent Flow Matching}
\name{Hsueh-Wei Fu, Yen-Tung Yeh, Yi-Hsuan Yang}
\address{National Taiwan University}
\begin{document}
%
\maketitle
\begin{abstract}

Realistic guitar synthesis requires both audio fidelity and interpretable control over \textit{how} notes are performed. The input timed MIDI events decide \textit{what} is played but not \textit{how}, leaving the continuous within-note decisions of guitar performance---such as pitch inflection and 
per-string articulation---unspecified. A model that directly maps MIDI to audio must therefore realize the performance implicitly, and users are left with limited control over the synthesized performance.
We therefore propose a two-stage guitar synthesis pipeline that predicts editable, per-string $f_0$ and amplitude trajectories from timed note events and note-level dynamics, and renders them into audio using latent flow matching. Experimental results show that the proposed system outperforms a matched single-stage counterpart and DDSP-Guitar in control fidelity and perceived quality, while keeping the controls editable. Sound examples can be found online~\footnote{https://weifu3024.github.io/ctrl-lfm-guitar-synth/}.

\end{abstract}


\section{Introduction}
\label{sec:intro}


Neural guitar synthesis, a subtask of MIDI-to-audio synthesis~\cite{wang2019performancenet,dong2022deepperformer,hawthorne2022spectrogram_diffusion,tang2025midivalle, wu2022midi_ddsp}, aims to render realistic musical performances from symbolic input such as MIDI or other score-level representations~\cite{jonason2024ddsp_guitar,wiggins2023differentiable_guitar,kim2024guitarroll}. Such systems lower the barrier to music production, allowing a composer to hear a human-like rendition of a score without hiring a guitarist.

Existing MIDI-to-audio approaches differ in the granularity of control they
expose to the neural networks for outputing audio. At the coarsest granularity, systems conditioned on MIDI alone~\cite{wang2019performancenet,dong2022deepperformer,hawthorne2022spectrogram_diffusion} receive only note-level information such as onsets, offsets, and pitches. For guitar, this leaves unspecified how each note is realized---the continuous pitch and amplitude trajectories produced on each string---which carries much of the instrument's expressiveness. Moreover, these MIDI-only approaches let the model learn this realization internally, leaving the user no way to adjust it.


Finer-grained approaches instead expose the performance itself. Recently, VIOLET~\cite{tian2026violet} has addressed this ambiguity in violin
synthesis by supplying detailed playing-technique labels directly to the neural network. Though effective, such technique labels demand expert knowledge of the instrument at inference time and are rarely annotated in the real recordings
required for training. A natural alternative is therefore to predict the controls: ViolinDiff~\cite{kim2025violindiff} predicts note-wise pitch bend before diffusion-based synthesis, while DDSP-based systems such as MIDI-DDSP~\cite{wu2022midi_ddsp} and DDSP-Guitar~\cite{jonason2024ddsp_guitar} predict intermediate acoustic features to retain controllability. Even with predicted controls, however, the realism of the synthesized audio still leaves room for improvement.

To preserve controllability while improving synthesis realism, we propose a two-stage framework in which a performance predictor infers per-string $f_0$ and amplitude trajectories from symbolic input (Sec.~\ref{ssec:predictor}, Figure~\ref{fig:perf-pred}), and a latent flow-matching~\cite{lipman2023flow} synthesizer renders them into audio (Sec.~\ref{ssec:synthesizer}, Figure~\ref{fig:lfm-synth}). We further introduce two mechanisms. First, to account for the strings of a guitar sounding simultaneously and coupling through the instrument body~\cite{debut2020physical}, we apply \textit{gated cross-string attention} in both stages, whereas DDSP-Guitar~\cite{jonason2024ddsp_guitar} uses it only in the prediction stage and processes strings independently in its synthesizer for its two stage system. Second, we propose a \textit{pseudo-velocity predictor} (Sec.~\ref{ssec:velocity}): unlike piano systems, where note velocity is a natural symbolic input~\cite{dong2022deepperformer,tang2025midivalle,shi2023tts_midi_audio}, guitar datasets derive it from the ground-truth audio~\cite{jonason2024ddsp_guitar,wiggins2023differentiable_guitar}, and adjusting it note by note is burdensome for the user. At inference, a user may therefore supply timed note events alone, or edit the predicted trajectories before rendering. Our experiments show that \textit{gated cross-string attention} benefits synthesis in polyphonic regions, and the \textit{pseudo-velocity predictor} enables MIDI-only synthesis while maintaining lower pitch and envelope errors than both baselines.

Our contributions are summarized as follows:
\begin{itemize}
    \item A two-stage controllable guitar synthesis framework that exposes
    editable string-wise $f_0$ and amplitude trajectories and renders them with
    a latent flow-matching synthesizer.

    \item \textit{Gated cross-string attention} in both stages, with an
    ablation that localizes where inter-string interaction matters.

    \item A \textit{pseudo-velocity predictor} that removes the dependence on
    velocity annotation, enabling MIDI-only synthesis.
\end{itemize}

\section{Proposed Method}
\label{sec:method}


We propose a two-stage pipeline for guitar synthesis. Let
$m \in \mathbb{R}^{4 \times S \times T}$ denote the frame-wise symbolic representation of the input MIDI, whose four channels are pitch $p$, voiced state $s$, relative note position $r$, and onset $o$, and let $v \in \mathbb{R}^{S \times T}$ denote the per-note velocity, where $S = 6$ is the number of strings and $T$ the number of frames. In the first stage, a performance predictor $\mathcal{P}$ maps them to the intermediate performance controls
$\mathcal{C} = (f_0, e_{\mathrm{rms}})$, where $f_0, e_{\mathrm{rms}} \in \mathbb{R}^{S \times T}$ are the per-string fundamental frequency and amplitude envelope trajectories. In the second stage, a latent flow-matching synthesizer $\mathcal{G}$ encodes $\hat{\mathcal{C}}$ and the symbolic channels of $m$, then uses an AdaLN-modulated DiT to generate per-string Stable Audio Open~\cite{evans2025stable_audio_open} latents. The SAO decoder converts them into per-string audio $\{x_i\}_{i=1}^{S}=\mathcal{G}(m,\hat{\mathcal{C}})$. Each $x_i$ is filtered by a string-specific impulse response, and the filtered signals are summed to produce $x_{\mathrm{mix}}$.


\subsection{Gated Cross-String Performance Predictor}
\label{ssec:predictor}
\begin{figure}[ht]
    \centering
    \includegraphics[width=\columnwidth]{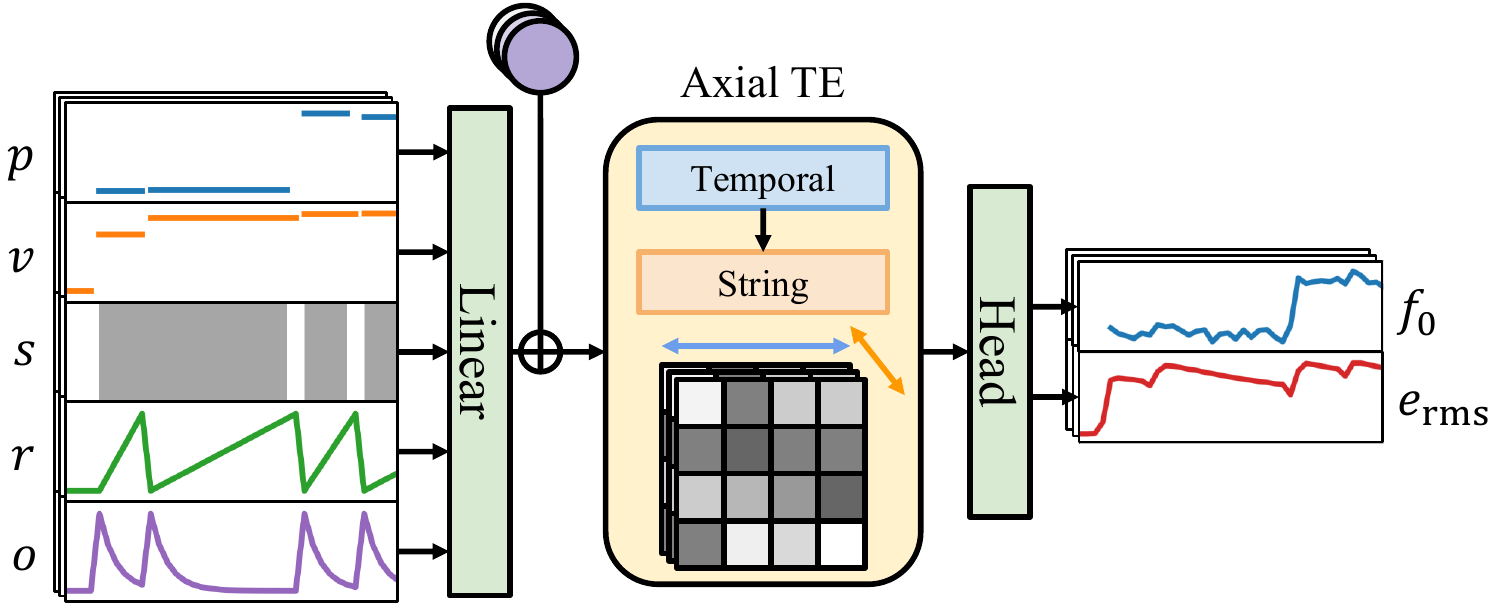}
    \caption{The architecture of the performance predictor. The purple circles represent the learned string identity embeddings.}
    \label{fig:perf-pred}
\end{figure}

The architecture of $\mathcal{P}$ is shown in Fig.~\ref{fig:perf-pred}. It takes the four channels of $m$ and the separate velocity channel $v$ as five frame-wise input features, each in $\mathbb{R}^{S \times T}$. The voiced state $s$, relative note position $r$, and onset $o$ are used to provide local note context: whether it lies inside a note, near the note's beginning or end, and close to an onset. A linear layer projects the five input features frame-wise to the model dimension, and a learned string-identity embedding is added to each string's channel, giving tokens localized in both time and string. 

Modeling these tokens requires two axes: a \emph{time} axis, along which each string's trajectory evolves within and across notes, and a \emph{string} axis, along which simultaneously sounding strings interact. Prior work models the time axis recurrently---a GRU over a context channel of temporally nearby notes~\cite{wiggins2023differentiable_guitar}, or an in-string LSTM~\cite{jonason2024ddsp_guitar}---while DDSP-Guitar’s two-stage models use cross-string attention in the control predictor but not in the synthesis decoder. In this work, we model both with an axial transformer~\cite{ho2019axial}: self-attention along the time axis, and \textit{gated cross-string attention} along the string axis. Each cross-string block is scaled by a zero-initialized gate, so training begins from purely per-string processing and cross-string information is introduced only as it proves useful; the learned gate magnitude then indicates how much of it the stage relies on. A final layer normalization and linear head map the encoder output to the per-string trajectories $f_0, e_{\mathrm{rms}} \in \mathbb{R}^{S \times T}$, which together form the controls $\mathcal{C}$.

\begin{figure}[ht]
    \centering
    \includegraphics[width=0.95\columnwidth]{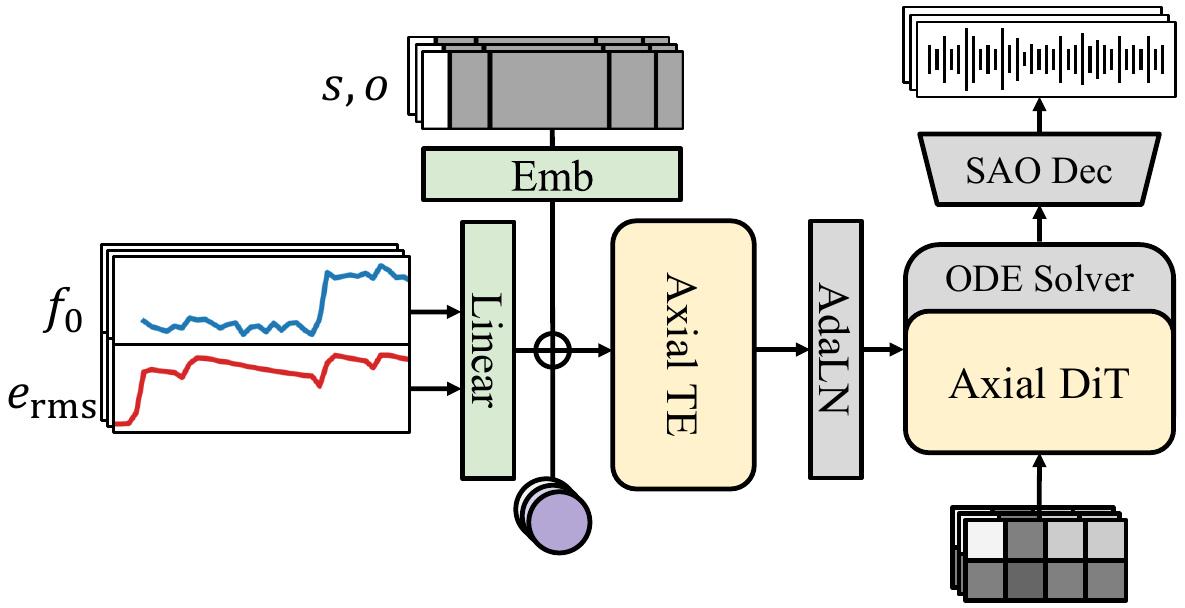}
    \caption{The architecture of the synthesizer. Each output single string audio will then be mixed by learned string-specific impulse responses.}
    \label{fig:lfm-synth}
\end{figure}

\subsection{Gated Cross-String Synthesizer}
\label{ssec:synthesizer}
In the second stage, the architecture of $\mathcal{G}$ is shown in Fig.~\ref{fig:lfm-synth}. It comprises a conditioning encoder and an axial DiT, connected through AdaLN modulation. In the conditioning encoder, a linear layer
projects the predicted controls $\hat{\mathcal{C}}$ to the model dimension, and a learned string-identity embedding is added. To mitigate the effect of imperfect predictions from the first stage, we explicitly re-use the voiced state $s$ and onset $o$ extracted from the MIDI, enabling the model to re-interpret unreliable regions: a learned embedding is added wherever a voiced or onset event occurs at a given frame and string. These tokens are then mixed by axial transformer blocks, and the resulting
representation modulates the axial DiT through AdaLN. The axial DiT, composed of
axial attention layers, predicts the vector field $u_\theta$, from which the
per-string SAO latents $z = \{z_i\}_{i=1}^{S}$ are solved with an Euler ODE solver, decoded by the pretrained SAO decoder into per-string audio $x_i$,
filtered by string-specific impulse responses, and summed into the final mixture $x_{\mathrm{mix}}$.



\subsection{Pseudo Velocity Predictor}
\label{ssec:velocity}
To demonstrate that our system can work without ground truth velocity, we also train a pseudo velocity predictor using only the MIDI information without velocity. The model encodes four note-level features—MIDI pitch, log-scaled duration, preceding inter-onset interval (IOI), and succeeding IOI—using a linear projection, and adds a learned string embedding to each note representation. The resulting sequence is processed by a transformer encoder, followed by a layernorm and linear head that predicts the pseudo velocity of each note. 

\subsection{Implementation Details}
For performance feature extraction, $f_0$ is extracted using CREPE \cite{kim2018crepe} over the range of 50–1100 Hz, and $e_{\text{rms}}$ is computed by taking the log-scale RMS curve in waveform domain. Both $f_0$ and $e_{\text{rms}} $ are computed on the same temporal grid as the Stable Audio Open latent representation, with a hop size of 2048 samples (21.53 Hz). Both features are standardized using statistics computed from the training split. To obtain the ground truth $v$ label, we follow the per-note peak-loudness conditioning strategy used in DDSP-Guitar and derive the pseudo-velocity from the peak value of our envelope representation within each note interval.

For the performance predictor, we use a model dimension of 256 and 6 axial Transformer encoder layers with 8 attention heads. The model is trained with frame-wise L1 loss $\mathcal{L}_{\mathrm{L1}}$, frame-wise first-difference L1 loss $\mathcal{L}_{\Delta}$ for trajectory supervision, and note-wise L1 loss $\mathcal{L}_{\mathrm{note}}$, where $ \mathcal{L}_{\mathrm{note}} $ first averages the frame-wise loss within each note and then averages across notes, balancing the contributions of notes with different durations. We use $\mathcal{L}=\mathcal{L}_{\mathrm{L1}}+\mathcal{L}_{\Delta}+1.3\mathcal{L}_{\mathrm{note}}$ and train for 150 epochs with a batch size of 4. For the synthesizer, the control encoder uses 3 layers, 4 attention heads, and a model dimension of 256, while the DiT uses 8 layers, 6 attention heads, and a model dimension of 384. The model is trained for 500 epochs with a batch size of 8 using MSE supervision on the rectified-flow velocity objective \cite{liu2023rectifiedflow} with logit-normal timestep sampling. To improve control adherence, we apply classifier-free guidance (CFG) \cite{ho2021classifierfree} with a conditioning dropout probability of 0.1 during training, and use a CFG scale of 2.0 with 50 Euler integration steps during inference. For the velocity predictor, we use 2 Transformer encoder layers, 4 attention heads, and a model dimension of 64. It is trained for 150 epochs with a batch size of 8 supervised on both L1 loss $\mathcal{L}_{\mathrm{L1}}$ and first-difference L1 loss $\mathcal{L}_{\Delta}$ computed between consecutive notes on the same string, i.e. $\mathcal{L}=\mathcal{L}_{\mathrm{L1}}+\mathcal{L}_{\Delta}$. All models are optimized using AdamW with a learning rate of 3e-4, a weight decay of 0.01, and gradient clipping at 1.0. 
For the mixing impulse responses, we train a linear FIR filter for each string (0.25 s at 44.1 kHz) such that the filtered per-string recordings sum to reconstruct the corresponding microphone recording. The filters are optimized for 200 epochs using a multi-scale spectral loss (MSSL) \cite{engel2020ddsp}, with a batch size of 4 and a learning rate of 2e-4.

\section{Experiments}
\label{sec:experiment}
\subsection{Experimental Setup}
\label{ssec:experimental-setup}

\noindent \textbf{Dataset.} We conduct experiments on GuitarSet~\cite{xi2018guitarset}, using the recording-level train/validation/test split adopted in DDSP-Guitar \footnote{https://github.com/erl-j/ddsp-guitar}, consisting of 306, 18, and 36 recordings, respectively. All six string recordings from the same performance are assigned to the same split. Audio is processed at 44.1 kHz, and training examples are randomly cropped to 128–256 frames, corresponding to approximately 5.9–11.9 s.

\noindent \textbf{Baselines.} We compare against two baselines.
\textit{Single-stage} is a variant of our model in which the synthesizer is conditioned directly on the nominal MIDI pitch and pseudo-velocity instead of the predicted $f_0$ and envelope, during both training and inference; it isolates the contribution of the intermediate control-prediction stage. \textit{DDSP-Guitar}~\cite{jonason2024ddsp_guitar} addresses string-wise guitar
synthesis on GuitarSet and provides a publicly available pretrained model\footnote{https://huggingface.co/erl-j/ddsp-guitar-unified}, enabling direct comparison on the same test split. We additionally report \textit{Oracle Synth}, the output of our synthesizer conditioned on the ground-truth controls, which serves as a reference upper bound on how well the
LFM synthesizer can follow its conditioning.

\begin{table*}[t]
\centering
\caption{Objective and subjective results. Ref. denotes method-specific audio-derived velocity; Pred. denotes MIDI-only predicted velocity. MOS compares ground truth with the three Ref. systems. Oracle synth uses reference controls. Dashes indicate inapplicable or unevaluated results.}
\label{tab:objective_results_merge}
\begingroup
\setlength{\tabcolsep}{3.5pt}
\renewcommand{\arraystretch}{0.9}
\setlength{\aboverulesep}{1pt}
\setlength{\belowrulesep}{1pt}
\begin{tabular}{@{}llccccccccc@{}}
\toprule
& & \multicolumn{4}{c}{Control fidelity}
& Audio & \multicolumn{4}{c}{MOS} \\
\cmidrule(lr){3-6} \cmidrule(lr){7-7} \cmidrule(lr){8-11}
System & Vel.
& $f_0$ MAE $\downarrow$ & $f_0$ Med. $\downarrow$
& Env. MAE $\downarrow$ & Env. Corr. $\uparrow$
& MSSL $\downarrow$
& CC $\uparrow$ & EX $\uparrow$ & AQ $\uparrow$ & OP $\uparrow$ \\
\midrule
\multicolumn{11}{@{}l}{\textit{References}} \\
\quad Ground truth & --
& -- & -- & -- & -- & --
& 3.94 & 3.92 & 3.84 & 3.91 \\
\quad Oracle synth & --
& 15.32 & 9.76 & 0.10 & 1.00 & 8.21
& -- & -- & -- & -- \\
\midrule
\multicolumn{11}{@{}l}{\textit{Baselines}} \\
\quad Single-stage & Ref.
& 33.53 & 11.47 & 0.80 & 0.63 & 9.43
& 2.85 & 2.92 & 2.88 & 2.54 \\
\quad DDSP-Guitar & Ref.
& 50.35 & 26.33 & 1.18 & 0.44 & \textbf{8.27}
& 3.09 & 2.69 & 2.92 & 2.62 \\
\midrule
\multicolumn{11}{@{}l}{\textit{Ours}} \\
\quad w/ oracle velocity & Ref.
& 26.52 & 10.46 & \textbf{0.30} & \textbf{0.88} & 8.79
& \textbf{3.46} & \textbf{3.24} & \textbf{3.10} & \textbf{3.04} \\
\quad w/ predicted velocity & Pred.
& \textbf{25.66} & \textbf{10.45} & 0.61 & 0.63 & 9.01
& -- & -- & -- & -- \\
\bottomrule
\end{tabular}
\endgroup
\end{table*}

\begin{table}[t]
\centering
\renewcommand{\arraystretch}{0.9}     
\setlength{\aboverulesep}{1pt}         
\setlength{\belowrulesep}{1pt}         
\caption{Cross-string attention ablation under oracle controls. \(\mathcal{G}_{\mathrm{w/o\,dec}}\) removes cross-string attention from the DiT; \(\mathcal{G}_{\mathrm{w/o\,all}}\) removes all cross-string attention from the synthesizer. Window-level \(\Delta\)MSSL is measured relative to the full model \(\mathcal{G}_{\mathrm{full}}\).}

\label{tab:cross_string_ablation}
\resizebox{\columnwidth}{!}{
\begin{tabular}{lccc}
\toprule
& Overall
& \multicolumn{2}{c}{Window-level $\Delta$MSSL} \\
\cmidrule(lr){2-2}
\cmidrule(lr){3-4}
System
& MSSL $\downarrow$
& Multi-string $\downarrow$
& Single-string $\downarrow$ \\
\midrule
$\mathcal{G}_{\text{full}}$
& \textbf{8.213}
& --
& -- \\

$\mathcal{G}_{\text{w/o dec}}$
& 8.363
& $+0.255$
& $+0.055$ \\

$\mathcal{G}_{\text{w/o all}}$
& 8.525
& $+0.614$
& $+0.158$ \\
\bottomrule
\end{tabular}
}
\end{table}

\subsection{Evaluation} 

\noindent\textbf{Objective Evaluation.} We report two groups of objective metrics. \textit{Control fidelity} is measured by the mean and median absolute $f_0$ error in cents over voiced frames ($f_0$ MAE, $f_0$ Med.), and by the MAE and Pearson correlation between envelopes (Env. MAE, Env. Corr.). \textit{Audio quality} is measured by the multi-scale spectral loss (MSSL)~\cite{engel2020ddsp}, following the configuration
of DDSP-Guitar, computed at 48\,kHz over the band below 22.05\,kHz shared by both systems. $f_0$ and envelope are re-extracted from the synthesized and ground-truth single-string signals using the procedure in Sec.~\ref{ssec:experimental-setup}, with all outputs resampled to 44.1\,kHz. 

\noindent \textbf{Subjective Evaluation. } As control and spectral metrics do not fully reflect perceptual quality, we also conduct an online listening test with 32 participants, rating each example on a 1--5 scale. Listeners rate four criteria: content correctness (CC), how faithfully the output follows the notes and rhythm of the score; expressiveness (EX), how natural and human-like the performance nuances are; audio quality (AQ), the overall fidelity of the output, independent of its content; and overall preference (OP). 
We select 18 eight-second excerpts from the GuitarSet test split and divide them into three disjoint sets of 6, with each participant evaluating one set.
In each trial, listeners first hear an unrated constant-velocity FluidSynth~\cite{FluidSynth} rendering of the MIDI as a content reference, then rate the four anonymized outputs in random order. 

\section{Results}

\noindent \textbf{Main results.} As shown in Table~\ref{tab:objective_results_merge}, our system outperforms both baselines in control fidelity. Compared with Single-stage, the oracle-velocity variant reduces the $f_0$ MAE from 33.53 to 26.52 cents and raises the envelope correlation from 0.63 to 0.88. We attribute this gap to the single-stage model having to infer performance with coherent pitch deviations, dynamics, and articulation implicitly while also learning audio generation, whereas our intermediate curves provide a temporally aligned performance specification that lets the synthesizer focus on mapping controls to audio. In terms of audio quality, MSSL is the only metric on which a baseline leads: DDSP-Guitar obtains a lower MSSL (8.27 vs.\ 8.79), as it is optimized directly on the mixed recording, whereas our synthesizer is supervised on isolated strings. This improves control fidelity but misses the recording-specific mixing characteristics that MSSL rewards. As for the system variant with predicted velocity, replacing oracle velocity with predicted velocity leaves pitch fidelity essentially unchanged ($f_0$ Med.\ 10.45 vs.\ 10.46 cents) and mainly affects the envelope, as velocity primarily governs note-level dynamics. Even with a slight performance drop, the predicted velocity variant still achieves higher control fidelity compared to the two baselines, showcasing our model's ability to synthesize from MIDI alone.

In the listening test (MOS columns in Table~\ref{tab:objective_results_merge}), our system with oracle velocity achieves the highest scores among synthesized systems on all four criteria (CC 3.46, EX 3.24, AQ 3.10, OP 3.04). The gains in CC and EX support explicitly modeling intermediate performance controls, and the higher AQ and OP over DDSP-Guitar, despite its lower MSSL, indicate that spectral distance does not fully capture perceived quality. Conversely, Single-stage outperforms DDSP-Guitar on control metrics yet receives a lower CC score (2.85 vs.\ 3.09), suggesting that frame-level metrics might miss perceptually salient errors such as missing notes, rhythmic inaccuracies, or localized artifacts.

\noindent \textbf{Cross String Attention Ablation.} We further conduct an ablation study to the gated-cross attention mechanism. Since the cross-string attention gates are zero-initialized, their learned values reveal whether cross-string information is used: they stay near zero in the performance predictor ($[-0.04, 0.05]$) but grow in the synthesizer's
control encoder ($[-0.17, 0.37]$) and DiT decoder ($[-0.40, 0.37]$). We therefore train two synthesizer variants, removing cross-string attention from the decoder ($\mathcal{G}_{\text{w/o dec}}$) or from the entire synthesizer ($\mathcal{G}_{\text{w/o all}}$), and compare them with the full model $\mathcal{G}_{\text{full}}$ under oracle controls, using
MSSL computed separately on single- and multi-string time windows. As shown in Table~\ref{tab:cross_string_ablation}, MSSL degrades progressively as more cross-string attention is removed, most strongly in multi-string windows, indicating that cross-string attention mainly benefits polyphonic regions.

\section{Conclusion}
\label{sec:conclusion}
In this work, we presented a controllable two-stage pipeline for guitar synthesis. Compared with a single-stage system, our approach improves per-string control fidelity and synthesis quality while providing editable intermediate pitch and envelope controls. We also introduced a pseudo-velocity predictor to enable synthesis when note-level velocity is unavailable. Our cross-string attention ablation shows that information exchange between strings is particularly beneficial in polyphonic regions. Future work will explore conditioning on timbre and recording environment, steering playing techniques through intermediate controls, and using string separation to reduce reliance on hexaphonic recordings.

\section{Acknowledgement}
The work is supported by grants the National Science and Technology Council of Taiwan (NSTC 114-2628-E-002-013-MY3), and the Ministry of Education (MOE) of Taiwan (for Taiwan Centers of Excellence in Artificial Intelligence). The authors declare no conflicts of interest. Codex, ChatGPT (OpenAI), Claude Code, Claude (Anthropic) assisted with code development and language editing.

\bibliographystyle{IEEEbib}
\bibliography{strings,refs}

\end{document}